\documentclass[12pt, amsfonts, amssymb]{article}

\usepackage{amssymb}
\usepackage{amsmath}
\usepackage{latexsym}

\numberwithin{equation}{section}

\newcommand{\op}{\oplus}
\newcommand{\CC}{{\mathbb C}}
\newcommand{\RR}{{\mathbb R}}
\newcommand{\ZZ}{{\mathbb Z}}

\newcommand{\Tr}{{\rm Tr}}

\newcommand{\ra}{\rightarrow}
\newcommand{\eps}{\epsilon}

\newcommand{\tA}{{\tilde A}}
\newcommand{\tF}{{\tilde F}}

\newcommand{\hG}{{\hat G}}
\newcommand{\hg}{{\hat g}}
\renewcommand{\O}{{\cal O}}
\newcommand{\tO}{{\tilde O}}
\newcommand{\htau}{{\hat\tau}}
\newcommand{\ot}{\otimes}
\newcommand{\tT}{{\tilde T}}
\newcommand{\ts}{{\tilde s}}
\newcommand{\tphi}{{\tilde\phi}}
\renewcommand{\Im}{{\rm Im}}
\renewcommand{\Re}{{\rm Re}}
\newcommand{\tG}{{\tilde G}}
\renewcommand{\t}{{\mathfrak t}}
\newcommand{\g}{{\mathfrak g}}
\renewcommand{\hg}{{\hat\g}}
\newcommand{\hatt}{{\hat{\mathfrak t}}}
\newcommand{\hPhi}{{\hat\Phi}}
\newcommand{\W}{{\mathcal W}}
\renewcommand{\sl}{{\mathsf{sl}}}
\newcommand{\so}{{\mathsf{so}}}
\renewcommand{\sp}{{\mathsf{sp}}}

\title{\bf \Large Wilson-'t Hooft operators in four-dimensional gauge theories and S-duality}
\author{Anton Kapustin\thanks{kapustin@theory.caltech.edu}\\
\it California Institute of Technology, Pasadena, CA 91125, USA
}

\begin{document}
\begin{titlepage}

\renewcommand{\thepage}{ }

\maketitle

\begin{abstract}

We study operators in four-dimensional gauge theories which are
localized on a straight line, create electric and magnetic flux,
and in the UV limit break the conformal invariance in the minimal
possible way. We call them Wilson-'t Hooft operators, since in the
purely electric case they reduce to the well-known Wilson loops,
while in general they may carry 't Hooft magnetic flux. We show
that to any such operator one can associate a maximally symmetric
boundary condition for gauge fields on $AdS_E^2\times S^2$. We
show that Wilson-'t Hooft operators are classifed by a pair of
weights (electric and magnetic) for the gauge group and its
magnetic dual, modulo the action of the Weyl group. If the
magnetic weight does not belong to the coroot lattice of the gauge
group, the corresponding operator is topologically nontrivial
(carries nonvanishing 't Hooft magnetic flux). We explain how the
spectrum of Wilson-'t Hooft operators transforms under the shift
of the $\theta$-angle by $2\pi$. We show that, depending on the
gauge group, either $SL(2,\ZZ)$ or one of its congruence subgroups
acts in a natural way on the set of
Wilson-'t Hooft operators. This can be regarded as evidence for
the S-duality of $N=4$ super-Yang-Mills theory. We also compute
the one-point function of the stress-energy tensor in the presence
of a Wilson-'t Hooft operator at weak coupling.

\end{abstract}

\vspace{-7in}

\parbox{\linewidth}
{\small\hfill CALT-68-2536}

\end{titlepage}
\pagestyle{empty}

\section{Introduction}

It is known to Quantum Field Theory aficionados, but not
appreciated widely enough, that local operators need not be
defined as local functions of the fields which are used to write
down the Lagrangian. The simplest example is an operator
which creates a winding state in the theory of a free periodic
boson in 2d. Another example is a twist operator for a single
Mayorana fermion in 2d. The second example shows that it is a
matter of convention which field is regarded as fundamental: the
massless Mayorana fermion can be reinterpreted as the continuum
limit of the Ising model at the critical temperature, and then it
is more natural to regard one of the twist operators (the one with
fermion number $0$) as the fundamental object, since it
corresponds to the spin operator of the Ising model. Another
famous example of this phenomenon is the fermion-boson equivalence
in two dimensions. The unifying theme of all these examples that
one can define a local operator by requiring the ``fundamental''
fields in the path integral to have a singularity of a prescribed
kind at the insertion point. It may happen that the presence of
the singularity can be detected from afar for topological reasons.
For example, in the presence of the fermionic twist operator, the
fermion field changes sign as one goes around the insertion point.
In such cases, one can say that the operator insertion creates
topological disorder. (In fact, all examples mentioned above
fall into this category). But it is important to realize that the
idea of defining local operators by means of singularities in the
``fundamental'' fields is more general than that of a topological
disorder operator.

Local operators not expressible as local functions of the
``fundamental'' fields play an important role in various
dualities. For example, a certain winding-state
operator in the theory of a free periodic boson in 2d is a fermion
which satisfies the equations of motion of the massless Thirring
model. Thus the massless Thirring model is dual to a free theory.
Upon perturbing the Thirring model Lagrangian by a mass term, one
finds that it is dual to the sine-Gordon model~\cite{Coleman}. The
fermion number of the Thirring model corresponds to the soliton
charge of the sine-Gordon model.

So far all our examples have been two-dimensional. Already in one of the first papers on the
sine-Gordon-Thirring duality it was proposed that similar dualities may exist in higher-dimensional
fields theories~\cite{Mandelstam}. More specifically, S.~Mandelstam proposed that an abelian gauge
theory in 3d admitting Abrikosov-Nielsen-Olesen (ANO) vortices may have a dual description where
the ANO vortex state is created from vacuum by a fundamental field. Much later, such 3d dual pairs
have indeed been discovered~\cite{IS,deBHOO,deBHOOY,deBHOY,five}. 
Duality of this kind is known as 3d mirror symmetry.
The operator in the gauge theory creating the ANO vortex
has been constructed in Refs.~\cite{BKW1,BKW2}. It is a topological disorder operator in the sense that
the gauge field looks like the field of Dirac monopole near the insertion point. In other words,
the operator is (partially) characterized by the fact that the first Chern class of the gauge bundle
on any 2-sphere enclosing the insertion point is nontrivial. In the dual theory, the same operator
is a local gauge-invariant function of the fundamental fields.

Analogous dualities exist for certain nonabelian gauge theories in 3d.
These theories do not have any conserved topological charges and therefore do not have topological disorder operators.
Nevertheless, they admit local operators characterized by the fact that near the insertion
point the gauge field looks like a Goddard-Nuyts-Olive (GNO) monopole~\cite{GNO}.
Such operators have been studied
in Ref.~\cite{Bo}, where it was argued that 3d mirror symmetry maps them to ordinary local
operators whose VEVs parametrize the Higgs branch of the dual theory.

It is natural to try to extend these considerations to 4d gauge
theories, the primary objective being a better understanding of
various conjectural dualities. In 4d a punctured neighborhood of a
point is homotopic to $S^3$, and since vector bundles on $S^3$ do
not have interesting characteristic classes, we do not expect to
find any local topological disorder operators in such theories.
But as we know from 3d examples, this does not rule out the
possibility of a local operator which creates a singularity in the
fields at the insertion point.

A more serious problem is that we are mostly interested in
operators which can be studied at weak coupling. This means that
the singularity in the fields that we allow at the insertion point
must be compatible with the classical equations of motion. If we
want the operator to have a well-defined scaling dimension, we
also have to require that the allowed singularity be
scale-invariant. Together these two requirements are too strong to
allow nontrivial solutions in four-dimensional field theories. For
example, in a theory of a free scalar field an operator insertion
at the origin can be thought of as a local modification of the
Klein-Gordon equation of the form
$$
\Box\phi(x)=A(\phi(x),\partial)\delta^4(x),
$$
where $A$ is a polynomial in the field $\phi$ and derivatives.
However, scale-invariance requires $A$ to have dimension $-1$,
which is impossible.  Similarly, in four dimensions no local
modifications of the Maxwell equations which would preserve
scale-invariance are possible. (The situation in two- and
three-dimensional field theories is very different in this
respect.)

To circumvent this problem, we recall that in some sense the most basic gauge-invariant operators in a gauge theory are not local, but distributed along a line. These are the famous Wilson loop operators~\cite{Wilson}.
Their importance stems from the fact that they serve as order parameters for confinement: in the
confining phase the expectation value of a Wilson loop in a suitable representation exhibits area 
law~\cite{Wilson}. Similarly, to detect the Higgs phase, one makes use of the 't Hooft loop 
operators~\cite{tHooftconf}. 
More general phases
with ``oblique confinement'' and mixed Wilson-'t Hooft loop operators also exist.

For line operators, there is no conflict between scale invariance
and the equations of motion. Consider, for example, an operator
supported on a straight line in $\RR^4$. It is clear that if we
consider a singularity in the fields which corresponds to a Dirac
monopole (embedded in the nonabelian gauge group), it will satisfy
both requirements. More generally, one can consider singularities
which correspond to a dyon worldline.

Another nice feature of line operators is that in some cases they can be regarded as topological
disorder operators.
Indeed, the punctured neighborhood of a straight line is homotopic to $S^2$, and in a nonabelian gauge theory
with gauge group $G$ one may consider gauge bundles which have a nontrivial 't Hooft magnetic flux
(a class in $H^2(S^2,\pi_1(G))$) on $S^2$. By definition, these are 't Hooft loop operators.
However, even if the gauge group is simply-connected, one may still consider nontopological line operators which are defined by the requirement that near the insertion line the gauge field looks like the field of a GNO monopole. This is completely analogous to the situation in 3d~\cite{Bo}. We will continue to call such operators 't Hooft operators, even though they may carry trivial 't Hooft magnetic flux. Note also
that the line entering the definition of the Wilson and 't Hooft operators can be infinite without boundary.
This is why we prefer to use the terms ``Wilson operator'' and ``'t Hooft operator'' instead of the more common names ``Wilson loop operator'' and ``'t Hooft loop operator.''

In this paper we study general Wilson-'t Hooft operators in 4d
gauge theories using the approach of Refs.~\cite{BKW1,BKW2,Bo}. We
are especially interested in the action of various dualities on
these operators. In this paper we focus on a particular example:
S-duality of the $N=4$ super-Yang-Mills (SYM) theory. As usually formulated, this conjecture
says that the $N=4$ SYM theory with a simple gauge Lie algebra $\g$
and complexified coupling
$\tau=\frac{\theta}{2\pi}+\frac{4\pi}{g^2}$ is isomorphic to a similar theory with the
Langlands-dual gauge Lie algebra $\hg$ and coupling $\htau=-1/\tau$~\cite{MO,WO,Osborn}.
We will focus on a corollary of this conjecture, which says that for a given gauge Lie
algebra $\g$ the self-duality group of $N=4$ SYM is either $SL(2,\ZZ)$ or one of its congruence
subgroups $\Gamma_0(q)$ for $q=2,3$, 
depending on $\g$~\cite{GGPZ,doreyetal}. We will refer to this group as the S-duality group.
A natural question to ask is how the S-duality group acts on the Wilson loop
operator of the theory. For example, in the simply-laced case, where the S-duality group
is $SL(2,\ZZ)$, one can ask how the generator $S$ acts on a Wilson loop operator.
The usual answer that it is mapped to the
't Hooft loop operator is clearly inadequate, since Wilson loop
operators are parametrized by irreducible representations of $\g$,
while according to 't Hooft's original definition~\cite{tHooftconf} 't Hooft loop
operators are parametrized by elements of $\pi_1(\hG)=Z(G)$, where $G$ is
the gauge group. In this paper we
describe the action of the full S-duality group on Wilson-'t Hooft
operators. The existence of such an action can be regarded as evidence for the S-duality
conjecture.

One can characterize a local operator $\O$ in a conformal field
theory by the OPE of $\O$ with conserved currents, for example the
stress-energy tensor $T_{\mu\nu}$. We define similar quantities
for line operators and compute them at weak coupling. We observe
that they are sensitive to the $\theta$-angle of the theory. This
is a manifestation of the Witten effect~\cite{witteneffect}.

The outline of the paper is as follows. In section~\ref{sec:defs}
we define more precisely the class of line operators we are going
to be interested in, as well as some of their quantitative
characteristics. In Section~\ref{sec:free} we study line operators
in free 4d theories. In particular, we discuss the action of
$SL(2,\ZZ)$ transformations on general Wilson-'t Hooft operators
in the free Maxwell theory. In section~\ref{sec:nonabelian} we
classify Wilson-'t Hooft operators in nonabelian gauge theories
and determine the action of the S-duality group on them. We also compute
the expectation value of the
stress-energy tensor in the presence of a Wilson-'t Hooft operator to leading order
in the gauge coupling. In
Section~\ref{sec:summ} we summarize our results and propose
directions for future work. In the appendix we collect some standard facts about compact simple
Lie algebras.

\section{The definition of line operators}\label{sec:defs}

Our viewpoint is that a field theory is defined by specifying a flow from a UV fixed point.
The UV fixed point is a Conformal Field Theory (CFT), and all local operators should be defined
in this CFT. This approach is especially convenient for studying local topological disorder operators,
because local operators in a flat-space $d$-dimensional CFT are in one-to-one correspondence
with states in the same CFT on $S^{d-1}\times\RR$. This can be seen by performing a Weyl rescaling of the flat Euclidean metric by a factor $1/r^2$, where $r$ is the distance to the insertion point. After Weyl rescaling, the metric becomes the standard metric on $S^{d-1}\times \RR$, while the insertion point is at
infinity (in the infinite past, if we regard the coordinate $\log r$ parametrizing $\RR$ as the Euclidean
time).

Similarly, line operators should also be defined in the UV fixed point theory, so from now on we limit
ourselves to line operators in CFTs. The main requirement that we impose on a line operator is that
its insertion preserves
all the space-time symmetries preserved by the line, regarded as a geometric object. This condition is meant
to replace the usual requirement that local fields be (quasi)primaries of the conformal group.
Of course, a generic line breaks all conformal symmetry, so to get an interesting restriction we
will limit ourselves to straight lines (this implicitly includes circular lines, because they are
conformally equivalent to straight lines). As in the case of local operators, it is useful (although
not strictly necessary) to perform a Weyl rescaling of the metric so that the line is at infinite
distance. In cylindrical coordinates, the metric of $\RR^d$ has the form
$$
ds^2=dt^2+dr^2+r^2 d\Omega_{d-2}^2,
$$
where $d\Omega_{d-2}^2$ is the standard metric on a unit $d-2$-dimensional sphere,
so it is natural to rescale the metric by a factor $1/r^2$, getting
$$
d\ts^2=\frac{dt^2+dr^2}{r^2}+d\Omega_{d-2}^2.
$$
This is the product metric on $H^2\times S^{d-2}$, where $H^2$ is the Lobachevsky plane, i.e.
the upper half-plane of $\CC$ with the Poincar\'{e} metric. Another name for $H^2$ is Euclidean
$AdS^2$. The line is at the boundary of $H^2$. After this Weyl rescaling, it is clear that the
subgroup of the conformal group preserved by the line is $SL(2,\RR)\times SO(d-1)$.

In this picture, a straight line operator corresponds to a choice
of a boundary condition for the path-integral of the CFT on
$H^2\times S^{d-2}$. By assumption, we only consider boundary
conditions which preserve the full group of motions of this space.
Possible boundary conditions for fields on AdS space have been
extensively studied in connection with AdS/CFT correspondence, see
e.g. Refs.~\cite{ADS1,ADS2}, and we will implicitly make use of some of
these results later on.

The use of the Weyl rescaling is not mandatory: it is useful only because it makes
the action of $SL(2,\RR)\times SO(d-1)$ more obvious. Alternatively, 
one can simply excise the line from $\RR^d$
and allow for fields to have singularities on this line compatible with the required symmetries.
We will often switch between the flat-space picture and the $H^2\times S^{d-2}$ picture.
We will distinguish fields on $H^2\times S^{d-2}$ with a tilde; this is necessary, because fields
with nonzero scaling dimension transform nontrivially under Weyl
rescaling. Specifically, in going from $\RR^d$ to $H^2\times S^{d-2}$ a tensor field of scaling dimension
$p$ is multiplied by $r^p$.

Given a line operator $W$, one may consider Green's functions of local operators with an insertion of $W$.
We will be especially interested in the normalized 1-point function of the stress-energy tensor
in the presence of $W$. We will denote it
$$
\langle T_{\mu\nu}(x)\rangle_W=\frac{\langle W\ T_{\mu\nu}(x)\rangle}{\langle W\rangle}.
$$
We use normalized Green's functions because in most cases of interest $W$ needs multiplicative
renormalization, and in normalized Green's functions the corresponding arbitrariness in the definition of
$W$ cancels between the numerator and the denominator.
Assuming that $T_{\mu\nu}$ is symmetric and traceless, the form of this 1-point function is completely fixed by the $SL(2,\RR)\times SO(d-1)$ invariance.
For definiteness, from now on we will specialize to 4d CFTs. (There are also interesting line operators in 3d CFTs; we plan to discuss them elsewhere.) Then the 1-point function of $T_{\mu\nu}$ takes the form:
$$
\langle T_{00}(x) \rangle_W=\frac{h_W}{r^4},\quad \langle T_{ij}(x) \rangle_W=
-h_W\frac{\delta_{ij}-2n_i n_j}{r^4},\quad \langle T_{0j}(x) \rangle_W=0.
$$
Here we used the coordinates
$x=(x^0,x^1,x^2,x^3)=(x^0,\vec{x})$, where $x^0=t$ and
$r=|\vec{x}|.$ We also let $n_i=x^i/r$. The coefficient $h_W$ is a
number characterizing the line operator $W$. In some sense it is
analogous to the scaling dimension of a local primary operator, so
we will call $h_W$ the scaling weight of $W$. This analogy does
not go very far though, because $h_W$ does not seem to admit an
interpretation in terms of commutation relations of $W$ with
conserved quantities. Indeed, to compute the vacuum expectation
value of the commutator of $W$ with a conserved charge
corresponding to a conformal Killing vector field $\xi^\mu$, one
has to integrate
$$
\xi^\mu\langle T_{\mu\nu}(x)\rangle_W
$$
over the 2-sphere given by the equation $r=\eps$, and then take the limit $\eps\ra 0$. But it is easy to
see that for Killing vector fields corresponding to $SO(3)$ symmetry the integral vanishes identically,
while for conformal Killing vector fields corresponding to $SL(2,\RR)$ symmetry it diverges as $1/\eps$.
(There is also a divergence due to the infinite length of the straight line). We will also see that
the scaling weight need not be positive, or even real.

Note that the above result for the 1-point function of $T_{\mu\nu}$ becomes much more obvious if one
uses the $H^2\times S^2$ picture. There, the expectation value of the tensor
$\tT=\tT_{\mu\nu} dx^\mu\ot dx^{\nu}$ takes the form
$$
\langle \tT\rangle_W = h_W (ds^2(H^2)-ds^2(S^2))
$$
It is obvious that this is the most general traceless symmetric tensor on $H^2\times S^2$ which is
invariant under all isometries.

\section{Line operators in free 4d CFTs}\label{sec:free}

To illustrate our approach to line operators, in this section we will
consider some examples in free 4d CFTs. This will also serve as a preparation for studying line
operators in $N=4$ super-Yang-Mills, which is an exactly marginal deformation of a free theory.

\subsection{Free scalar}

We begin with the theory of a free scalar in 4d. The flat-space Euclidean action is
$$
S=\frac{1}{2g^2}\int d^4x\, (\partial\phi)^2.
$$
This action is invariant under Weyl rescaling if in curved space-time we allow for an extra term in
the action $\frac{1}{6} R\phi^2.$ The improved stress-energy tensor (which is also the Einstein
tensor for the action with the extra term) is
\begin{equation}\label{Tsc}
T_{\mu\nu}=g^{-2}\left[\partial_\mu\phi\,\partial_\nu\phi-\frac{1}{2}\delta_{\mu\nu}(\partial\phi)^2
-\frac{1}{6}\left(\partial_\mu\partial_\nu-\delta_{\mu\nu}\partial^2\right)\phi^2\right].
\end{equation}
This CFT admits a line operator
\begin{equation}\label{opscalar}
V_\lambda=\exp\left(\lambda\int\phi(t,\vec{0}) dt\right),\quad \lambda\in \CC.
\end{equation}
It is easy to see that it preserves the required symmetries (basically, this follows from the fact
that $\phi$ is a primary field of dimension $1$). Evaluating the 1-point function of $T_{\mu\nu}$
in the usual way, we find the scaling weight of $V_\lambda$:
\begin{equation}\label{hscalar}
h_{\lambda}=-\frac{\lambda^2 g^2}{96\pi^2}
\end{equation}

In the $H^2\times S^2$ picture, the boundary condition is the ``free'' boundary condition.
This means that in the limit $r\ra 0$ the field $\tphi=r\phi$ behaves as follows:
\begin{equation}\label{adsscfree}
\tphi=a(t)+\tO(r),
\end{equation}
where the function $a$ is not constrained. The symbol $\tO(r^n)$ means
``of order $r^n$ in the $H^2\times S^2$ picture,'' while $O(r^n)$ will mean ``of order $r^n$ in the $\RR^4$ picture.'' Later, when we consider tensor fields on $H^2\times S^2$ (resp. $\RR^4$), we will say
that a tensor is $\tO(r^n)$ (resp. $O(r^n)$) if its components in an orthonormal basis
are of order $r^n$.

If we use the $\RR^4$ picture, Eq.~(\ref{adsscfree}) means that $\phi$ is allowed
to have a singularity of the form
$$
\phi=\frac{a}{r}+O(1).
$$
In addition, the path-integral contains a factor $V_\lambda$ (in either picture). Note that this factor, as well
as the classical action, are UV divergent for $a\neq 0$. It is convenient to deal with this divergence by restricting the region of integration in the expresssion for the action to $r\geq \eps>0$ and by regularizing $V_\lambda$ as follows:
$$
V_\lambda(\eps)=\exp\left(\frac{1}{4\pi}\lambda\int\phi(t,\eps\vec{n})dt
d^2\sigma\right).
$$
Here $d^2\sigma$ is the area element of a unit 2-sphere
parametrized by $\theta,\phi$, and $\vec{n}$ denotes a unit vector
in $\RR^3$ pointing in the direction specified by $\theta,\phi$.
In the end we have to send $\eps$ to zero.

Since we are dealing with a Gaussian theory, the normalized
1-point function of $T_{\mu\nu}$ is simply the value of
$T_{\mu\nu}$ on the solution of classical equations of motion. The
only solution satisfying the required boundary condition at $r=0$
and preserving all the symmetries is
$$
\phi_{cl}=\frac{a_0}{r},\quad a_0\in\RR .
$$
Note that in the $H^2\times S^2$ picture this is simply a constant scalar field: $\tphi=a_0$.
It obviously preserves the full isometry group of $H^2\times S^2$, which is one way to explain
why $V_\lambda$ is a good line operator.

The constant $a_0$ is determined by requiring that the variation
of $\log V_\lambda$ to cancel the boundary term in the variation
of the classical action. The former is
$$
\frac{1}{4\pi\eps}\int \lambda \delta a dt d^2\sigma,
$$
while the latter is
$$
\frac{1}{\eps g^2} \int a\delta a dtd^2\sigma.
$$
Requiring the equality of the two terms, we find
$$
a_0=\frac{\lambda g^2}{4\pi}.
$$
Plugging this scalar background into the classical expression Eq.~(\ref{Tsc})
for $T_{\mu\nu}$, we find again the result Eq.~(\ref{hscalar}).

\subsection{Free Maxwell theory}

Next we consider the free Maxwell theory. The flat-space action is
$$
S_0=\frac{1}{4g^2} \int d^4 x\, F_{\mu\nu} F_{\mu\nu}.
$$
The form $A=A_\mu dx^\mu$ has scaling dimension $0$, so we do not have to make a distinction between
$A$ and $\tA$, or $F$ and $\tF$. The stress-energy tensor is
$$
T_{\mu\nu}=g^{-2}\left[-F_{\mu\lambda}F_{\nu\lambda}+\frac{1}{4}\delta_{\mu\nu} F_{\lambda\rho} F_{\lambda\rho}\right].
$$
The simplest line operator is the Wilson operator
$$
W_n=\exp\left(i n\int_L A_\mu dx^\mu\right).
$$
If the gauge group is compact, then $n$ must be an integer. The
operator $W_n$ inserts an infinitely massive particle of charge
$n$ whose worldline is $L$. An easy computation gives the scaling
weight of $W_n$:
\begin{equation}\label{hwilson}
h(W_n)=\frac{n^2g^2}{32\pi^2}.
\end{equation}
In the $H^2\times S^2$ picture, one considers the path integral
over topologically trivial gauge fields with free boundary
conditions for $F_{0r}$ and fixed boundary conditions for the rest
of the components of $F=dA$ at $r=0$:
$$
F=dA= (a(t)+O(r))\frac{dr\wedge dt}{r^2}+O(1).
$$
where the function $a(t)$ is arbitrary. In the gauge $A_r=0$, the
corresponding boundary condition on $A$ reads
$$
A=(a(t)+O(r))\frac{dt}{r}+O(1).
$$
One also needs to insert $W_n$ in the path-integral.

The corresponding solution of the classical equations of motion is
simply a constant electric field on $H^2\times S^2$:
$$
F=a_0\frac{dr\wedge dt}{r^2}.
$$
In the $\RR^4$ picture, this is simply the Coulomb field of a charged particle, as expected.
The magnitude of the electric field is again determined by requiring the cancellation of the
boundary variation of $S$ and the variation of $W_n$. Substituting this classical solution into the
classical $T_{\mu\nu}$, one again finds Eq.~(\ref{hwilson}). Note that a constant electric field
on $H^2$ is clearly invariant under isometries. This shows that the Wilson operator is a good line
operator in our sense.

By electric-magnetic duality, we expect to have a ``magnetic'' Wilson line parametrized by an integer
$m$. By definition, this is an 't Hooft operator corresponding to an insertion of a Dirac monopole of
charge $m$. In the $H^2\times S^2$ picture, this clearly means that $F$ has the following boundary behavior:
$$
F=\frac{m}{2} vol(S^2)+O(1).
$$
If the gauge group is compact, $m$ must be an integer. In the
gauge $A_r=0$, both $A_0$ and $A_i$ have fixed boundary
conditions:
$$
A_0=O(1),\quad A_i dx^i=\frac{m}{2}(1-\cos\theta)d\phi+O(1).
$$
The unique solution of equations of motion satisfying these boundary conditions is
$$
F=\frac{m}{2} vol(S^2),
$$
i.e. a constant magnetic field on $S^2$. It is obviously
invariant under all isometries of $H^2\times S^2$. From
the $\RR^4$ perspective, this is the field of a Dirac monopole of magnetic charge $m$:
$$
F=\frac{m}{4} \frac{\eps_{ijk} x^i dx^j\wedge dx^k}{r^3}.
$$
Substituting this solution into the classical $T_{\mu\nu}$, we
find the scaling weight of the 't Hooft operator:
\begin{equation}\label{hthooft}
h(H_m)=\frac{m^2}{8g^2}.
\end{equation}
This result is exact, because we are dealing with a Gaussian theory. Note that $h(W_n)$ at gauge coupling
$g^2$ is equal to $h(H_n)$ at gauge coupling ${\hat{g}}^2=\frac{4\pi^2}{g^2}$.
This follows from the S-duality of the Maxwell theory on $\RR^4$: the inversion of the gauge coupling
$$
\tau=\frac{2\pi i}{g^2}\mapsto \htau=-\frac{1}{\tau}
$$
leaves the theory invariant and maps $W_n$ to $H_n$.

Next we generalize this discussion in two directions. First of all, we will consider ``mixed''
Wilson-'t Hooft operators corresponding to the insertion of an infinitely massive dyon with an electric
charge $n$ and magnetic charge $m$. Second, we will allow for a nonvanishing $\theta$-angle.

We impose the following boundary conditions on the gauge fields on $H^2\times S^2$:
$$
F= (a(t)+O(r))\frac{dr\wedge dt}{r^2}+\frac{m}{2} vol(S^2)+O(1).
$$
That is, asymptotically we have a fixed magnetic field on $S^2$
and an electric field of an unconstrained magnitude. We also
insert a factor $W_n$ in the path-integral. The action is now
$$
S_\theta=S_0-\frac{i\theta}{8\pi^2}\int F\wedge F.
$$
Since the $\theta$-term is a total derivative, its variation is a boundary term, and therefore
affects the value of $a_0$. Indeed, the boundary variation of the full action is now
$$
\frac{4\pi}{g^2}\frac{1}{\eps}\int a\delta a dt
-\frac{im\theta}{2\pi}\frac{1}{\eps} \int \delta a dt
$$
while the variation of $W_n$ is the same as before:
$$
\delta \log W_n=in\frac{1}{\eps}\int \delta a dt.
$$
Requiring their equality, we get
$$
a_0=\frac{ig^2}{4\pi}\left(n+\frac{m\theta}{2\pi}\right).
$$
Evaluating $T_{\mu\nu}$ on the corresponding classical
background, we find the scaling weight of the Wilson-'t Hooft operator:
$$
h(WH_{(n,m)})=\frac{g^2}{32\pi^2}\left[\left(n+\frac{m\theta}{2\pi}\right)^2+\frac{4\pi^2m^2}{g^2}\right].
$$
Introducing the complexified gauge coupling
$$
\tau=\frac{2\pi i}{g^2}+\frac{\theta}{2\pi},
$$
we can rewrite this in a manifestly $SL(2,\ZZ)$-covariant way:
$$
h(WH_{(n,m)})=\frac{1}{16\pi}\frac{|n+m\tau|^2}{\Im\tau}.
$$
The S and T transformations act as follows:
\begin{align}
 & S: \tau\mapsto -1/\tau, & (n,m)\mapsto  (-m,n),\\
 & T: \tau\mapsto \tau+1, & (n,m)\mapsto (n+m,m).
\end{align}

\subsection{BPS Wilson-'t Hooft operators in $N=4$ Maxwell theory}

Let us now consider $N=4$ supersymmetric Maxwell theory. This is a product of the free Maxwell theory,
four copies of the free fermion theory, and six copies of the free scalar theory. In such a theory
it is natural to consider BPS line operators, i.e. line operators which preserve some supersymmetry.
We can take the following ansatz for such an operator:
$$
WH^{BPS}_{(n,m)}=W_n H_m V_\lambda,
$$
where $V_\lambda$ is the line operator Eq.~(\ref{opscalar}) for one of the six scalar fields.
The coefficient $\lambda$ is fixed by the BPS requirement. Note that it is incorrect, in general, to
fix $\lambda$ by requiring the expression
$$
W_n V_\lambda=\exp\left(\int\left(i n A_0+\lambda\phi\right) dt\right)
$$
to be preserved by some supersymmetries. This would ignore the
possibility that the SUSY variation of the action produces a
boundary term, which has to be cancelled by the SUSY variation of
$W_n V_\lambda$. To find the right value of $\lambda$, one can
either analyze in detail these boundary terms, or, more simply,
require that the corresponding solution of the classical equations
of motion be BPS. The relevant solution of the equations of motion
is
$$
F=\frac{i}{2}\frac{\Re (n+m\tau)}{\Im\tau}\frac{dt\wedge dr}{r^2}+\frac{m}{2} vol(S^2),\quad
\phi=\frac{\lambda g^2}{4\pi r}.
$$
This field configuration is a BPS dyon if and only if
$$
\lambda=|n+m\tau|.
$$
Then the scaling weight of the BPS Wilson-'t Hooft operator is
$$
h(WH^{BPS}_{(n,m)})=\frac{1}{24\pi}\frac{|n+m\tau|^2}{\Im\tau}.
$$
This expression is exact because the theory is Gaussian.

\section{Wilson-'t Hooft operators in nonabelian 4d gauge theories}\label{sec:nonabelian}

In this section we study Wilson-'t Hooft operators in conformally-invariant
nonabelian gauge theories in 4d. Some of our statements apply to any such theory, while others apply only
to the most well-known example, the $N=4$ super-Yang-Mills theory. The gauge group, to be denoted $G$,
is assumed to be simple and compact. Its Lie algebra will be denoted $\g$.

\subsection{Wilson operators}

Wilson operators in a theory with a gauge group $G$ are classified by irreducible representations
of $G$ and have the form
$$
W_R=\Tr_R\ P\exp\left(i\int A_0 dt\right),
$$
where $\Tr_R$ is the trace in the irreducible representation $R$.
In the $H^2\times S^2$ picture and in the gauge $A_r=0$, one
imposes the following boundary conditions:
$$
A_0=\frac{a(t)}{r}+O(1),\quad A_i=O(1),
$$
where $a(t)$ is an arbitrary function valued in the Lie algebra
$\g$. That is, the boundary conditions for $A_0$ are free, while
the boundary conditions for $A_i$ are fixed. One also has to
insert the operator $W_R$ in the path-integral.

Evaluating the expectation value of the stress-energy tensor in the presence of
the Wilson line at weak coupling, one easily finds the leading-order result for the scaling weight:
\begin{equation}\label{scweightW}
h\left(W_R\right)=\frac{g^2 c_2(R)}{64\pi^2}+O(g^4),
\end{equation}
where $c_2$ is the second Casimir of the representation $R$.

\subsection{'t Hooft operators and GNO monopoles}

Generalizing the results of Section~\ref{sec:free}, we expect that
't Hooft operators correspond to fixed boundary conditions for all
the fields. We also expect $A_0$ to be of order $O(1)$ at $r=0$.
Together with the requirement of $SL(2,\RR)\times SO(3)$
invariance, this completely fixes the boundary behavior of
$F=dA+A\wedge A$:
\begin{equation}\label{hooftbdrycond}
F=\frac{B}{2} vol(S^2)+O(1),
\end{equation}
where $B$ is a section of the adjoint bundle on $S^2$. The section
$B$ does not have to be constant (such a constraint would not be
gauge-invariant anyway), but if we want this asymptotics to respect
the $SO(3)$ isometry, then $B$ must be covariantly constant. It
was shown in Ref.~\cite{GNO} that this implies a quantization law
for $B$:
\begin{equation}\label{expcond}
\exp(2\pi i B)=id_G.
\end{equation}
One can always use gauge transformations to make $B$ at any chosen
point on $S^2$ to belong to some fixed Cartan subalgebra
$\t\subset\g$. If the adjoint bundle is trivial, we can regard $B$
as a constant valued in $\t$ and satisfying Eq.~(\ref{expcond}).
If the adjoint bundle is nontrivial, we can cover $S^2$ with two
charts (by removing either the south or the north pole), and in
each of the charts we can choose a trivialization where $B$ is a
$\t$-valued constant. In such a trivialization, the gauge field
asymptotics has the form
$$
A_\mu dx^\mu=\frac{B}{2} (1-\cos\theta) d\phi + O(1),
$$
where we only wrote down the expression in the chart covering the
north pole ($\theta=0$). This is simply a Dirac monopole embedded
into the nonabelian group $G$.

Note that Goddard et al. regarded Eq.~(\ref{hooftbdrycond}) as
describing the asymptotics of the gauge field at $r\ra\infty$,
while we regard it as describing the asymptotics at $r\ra 0$. Thus
although the mathematical manipulations are the same as in
Ref.~\cite{GNO}, the physical interpretation is different.

Let us first specialize to the case when the gauge group $G$ has a trivial center. That is,
$$
G=G_0:=\tG/Z(\tG),
$$
where $\tG$ is the unique simply connected compact Lie group with Lie algebra $\g$.
This case is of particular interest to us, because in the $N=4$ super-Yang-Mills theory all the
fields are in the adjoint representations of $\g$, and it is natural to take $G_0$
as the gauge group. Note that the action of $N=4$ super-Yang-Mills depends only on $\g$, not
on the Lie group $G$, thus the choice of $G$ is up to us. In flat space-time, the only effect of
taking a different $G$ with the same Lie algebra $\g$ would be to restrict the allowed values of the
't~Hooft magnetic flux for line operators. Since we would like to classify all possible
line operators, we choose not to put any unnecessary restrictions on the magnetic flux.

If $G=G_0,$ then the condition Eq.~(\ref{expcond}) is equivalent to the following quantization law~\cite{GNO}:
\begin{equation}\label{GNOquant}
\alpha(B)\in \ZZ,\quad\forall \alpha\in \Phi.
\end{equation}
Here $\Phi\subset \t^*$ is the set of roots of $\g$. Our Lie algebra conventions are summarized in
the Appendix. We find it convenient to keep the normalization of the Killing metric on $\g$ undetermined
until as late as possible. Therefore we do not identify $\t$ and $\t^*$, and regard roots and weights
as elements of $\t^*$, while coroots $H_\alpha$ are regarded as elements of $\t$.

The element $B\in\t$ is defined only modulo the action of the Weyl group~\cite{GNO}:
$$
w_\alpha: B\mapsto B'=B-\alpha(B) H_\alpha,\quad \alpha\in\Phi.
$$
Since $\beta(H_\alpha)$ is integer for any two roots $\alpha,\beta$, the quantization condition
is obviously invariant under the action of the Weyl group.

Let us pick an arbitrary $Ad$-invariant metric on $\g$. The Langlands-dual Lie algebra
$\hg$ is defined as follows. Its Cartan subalgebra is $\hatt=\t^*$, and the set
of roots $\hPhi\subset \t$ is defined to be the set of coroots of $\g$. Then the set of coroots of $\hg$
coincides with the set of roots of $\g$. The Weyl groups
of $\g$ and $\hg$ are the same. Then $B$ can be regarded as
an element of $\hatt^*$ which takes integer values on any coroot of $\hg$. This is precisely
the definition of a weight of $\hg$. Thus specifying the equivalence class of $B$ under the action of the
Weyl group which satisfies the quantization law Eq.~(\ref{GNOquant}) is the same as specifying
a dominant weight of $\hg$. The latter is the same as specifying an irreducible representation of $\hg$.
This observation was made for the first time in Ref.~\cite{GNO}, where for this reason elements
of $\t$ satisfying Eq.~(\ref{GNOquant}) were called magnetic weights. Magnetic weights form
a lattice in $\t$ which contains the lattice spanned by all coroots $H_\alpha$ (see Appendix).

We conclude that 't Hooft operators in a gauge theory with a
centerless simple compact Lie group $G$ are classified by
irreducible representations of $\hg$, or, equivalently, by orbits
of magnetic weights in $\t$ under the action of the Weyl group of
$\g$. We will denote the 't Hooft operator corresponding to a
magnetic weight $B$ by $H_B$. This is a finer classification than
the classification of 't Hooft operators by their 't Hooft
magnetic flux. Indeed, the latter takes values in
$\pi_1(G_0)=Z(\tG)$, which can be identified with the quotient
$$
\Lambda_{mw}/\Lambda_{cr}
$$
where $\Lambda_{mw}\in \t$ is lattice of magnetic weights, and
$\Lambda_{cr}\in \t$ is the coroot lattice. The 't Hooft magnetic
flux of a line operator $H_B$ can be identified with the image of
$B\in \Lambda_{mw}$ under the projection
$$
\Lambda_{mw}\ra \Lambda_{mw}/\Lambda_{cr}.
$$

If the gauge theory in question contains representations of $\g$ which transform nontrivially
under $Z(\tG)$ (e.g. if $\g=\sl_N$ and there are fields in the fundamental representation of $SU(N)$),
then the gauge group is some cover of $G_0$. Let $\Gamma\subset \t$ denote the kernel
of the exponential mapping
$$
\exp: \t\ra G,\quad B\mapsto \exp(2\pi i B).
$$
The set $\Gamma$ is a lattice in $\t$, and the condition Eq.~(\ref{expcond}) says that $B\in \Gamma$.
The lattice $\Gamma$ is contained in $\Lambda_{mw}$ and contains $\Lambda_{cr}$. The center of $G$
is actually the quotient $\Lambda_{mw}/\Gamma$, while the fundamental group $\pi_1(G)$ is the
quotient $\Gamma/\Lambda_{cr}$. Thus if the center of the gauge group $G$ is nontrivial, one gets
an extra constraint on the 't Hooft magnetic flux of 't Hooft operators: the flux must lie in
$\Gamma/\Lambda_{cr}$, which is a subgroup of $\Lambda_{mw}/\Lambda_{cr}.$ The lattice $\Gamma$
can also be thought of as the weight lattice of some compact Lie group $\hG$ with Lie algebra $\hg$.
Namely, $\hG$ is the group whose center is $\Gamma/\Lambda_{cr}$. This group was introduced
in Ref.~\cite{GNO}, where it was proposed that a gauge theory with gauge group $G$ may admit a dual
description as a gauge theory with gauge group $\hG$. For this reason is usually called the GNO-dual of $G$ in the physics literature. In the mathematical literature $\hG$ is called the Langlands-dual of $G$ because of its role in the Langlands program.

\subsection{Classification of Wilson-'t Hooft operators}

Now we consider mixed Wilson-'t Hooft operators. A fundamental
observation made in Refs.~\cite{Abou1,NM,Abou2} (see also
Ref.~\cite{HR}) is that in the presence of a nonabelian magnetic
field global gauge transformations are restricted to those which
preserve the Lie algebra element $B$. Thus we expect Wilson-'t
Hooft operators to be labelled by a pair $(B,R)$, where $B\in \t$
is a magnetic weight, and $R$ is an irreducible representation of
the stabilizer subgroup of $B$.

Before showing that this is indeed the case, we need to address
one possible source of confusion. When considering 't Hooft
operators, it was natural to take $G$ to have the smallest
possible center. In particular, for $N=4$ super-Yang-Mills it was
natural to take $G$ to have a trivial center. On the other hand,
when considering Wilson operators in the same theory, it is
natural to allow arbitrary irreducible representations of $\tG$,
the universal covering group of $G_0$. In general, a
representation of $\tG$ is only a projective representation of
$G_0$, so one may wonder if $R$ in the pair $(B,R)$ must be a
representation of the stabilizer of $B$ in $G_0$ or in $\tG$. The
latter is the correct answer. The simplest way to see this is to
note that in the case $B=0$ it gives the correct result (that
Wilson operators are classified by representations of $\tG$). Here
is another way to argue the same thing. Basically, we are saying
that if the dynamical fields all transform trivially under
$Z(\tG)$, then it makes sense to probe the theory with a source
which transforms in an arbitrary representation of $Z(\tG)$ and
carries an arbitrary 't Hooft magnetic flux in $\pi_1(G_0)$. This
is familiar from the abelian case: if there is a single dyon in
the universe, and no other electrically or magnetically charged
particles, then the Dirac-Zwanziger-Schwinger condition is
vacuous, and the electric and magnetic charges are completely
arbitrary. Nontrivial conditions arise only when we consider a
pair of dyons. In the present context, we expect a constraint on
pairs of allowed Wilson-'t Hooft operators arising from the
requirement of locality.

Coming back to the issue of classification of Wilson-'t Hooft
operators, we will impose the following (fixed) boundary
conditions on $A_i$:
$$
A_i dx^i=\frac{B}{2}(1-\cos\theta)d\phi+O(1),
$$
where $B$ is a covariantly-constant section of the adjoint bundle
on $S^2\times \RR$. This implies that the ``spatial'' components
of $F$ have the following boundary behavior:
$$
\frac{1}{2}F_{ij} dx^i dx^j=\frac{B}{4} \frac{\eps_{ijk} x^i dx^j\wedge dx^k}{r^3}+O(1).
$$
As for $A_0$, we attempt to impose the free boundary condition,
i.e.
$$
A_0=\frac{a}{r}+O(1),
$$
where $a$ is an arbitrary section of the adjoint bundle on
$S^2\times\RR$. We have to check whether these boundary conditions
preserve $SL(2,\RR)\times SO(3)$ invariance. The only nontrivial
constraint comes from requiring invariance with respect to
translations of $t=x^0$:
\begin{equation}\label{covconst}
D_0 F_{ij}=0.
\end{equation}
In the abelian case, this condition required $B$ to be
time-independent. In the present case, this condition first of all
implies that we can choose a local trivialization of the adjoint
bundle on $S^2\times\RR$ by two charts of the form
$U_\pm\times\RR$, where $U_+,U_-$ is the standard covering of
$S^2$, so that $B$ is independent of $t$. In such a
trivialization, the condition Eq.~(\ref{covconst}) becomes
$$
[A_0,B]=O(r).
$$
In particular, the section $a$ should commute with the section
$B$.

We have learned that in the neighborhood of $r=0$ the gauge field
$A_0$ takes value in the centralizer of $B$ in $\g$. We will
denote this centralizer $\g_B$. The gauge group $\tG$ is also
broken down to $\tG_B$, the stabilizer of $B$ in the adjoint
representation of $\tG$. The Lie algebra of $\tG_B$ is $\g_B$.
Thus we can construct a line operator by picking an irreducible
representation $R$ of $\tG_B$ and inserting in the path-integral
the factor
\begin{equation}\label{wilsonR}
W_R=\Tr_R P\exp\left(i\int A_0(t,0) dt\right).
\end{equation}
We conclude that Wilson-'t Hooft operators are classified by a
pair $(B,R)$, where $B$ is an equivalence class of a magnetic
weight, and $R$ is an irreducible representation of $\tG_B$, the
stabilizer of $B$.

In the next subsection we will discuss the action of the S-duality
group on the set of Wilson-'t Hooft operators. As a preparation,
let us repackage the data $(B,R)$ in a more suggestive form. The
group $\tG_B$ is a compact connected reductive Lie group whose Lie
algebra we have denoted $\g_B$. The Cartan subalgebra of $\g_B$
can be identified with $\t$. Let $\Phi_B$ be the set of roots of
$\g_B$. It is easy to see that $\Phi_B$ is a subset of $\Phi$
defined by the condition
\begin{equation}\label{phiBcond}
\alpha(B)=0.
\end{equation}
Further, the Weyl group $\W_B$ of $\g_B$ is a subgroup of the Weyl group $\W$ of $\g$ and
consists of reflections associated to roots in $\Phi_B$. In other words, $\W_B$
is the stabilizer subgroup of $B$ in $\W$. The maximal tori of $\tG_B$ and $\tG$ coincide, so
a representation of $\tG_B$ can be specified by specifying how the maximal torus of $\tG$ acts.
This can be done by picking a weight of $\g$. If we were discussing
representations of $\tG$, we would also identify weights lying in the same $\W$-orbit. But since we
are interested in representations of $\tG_B$, we only need to identify weights lying in the same
$\W_B$ orbit.

To summarize, specifying a $\W$-orbit of a magnetic weight
$B\in\t$ and an irreducible representation $R$ of $\tG_B$ is the
same as specifying a $\W$-orbit of $B\in\t$ and a $\W_B$ orbit of
an ordinary (``electric'') weight $\mu\in \t^*$. Here $\W_B$ is
the subgroup of $\W$ leaving $B$ invariant. But clearly this is
the same as specifying a pair of weights, one magnetic and one
electric, and identifying pairs related by the action of $\W$.
Thus Wilson-'t Hooft operators are classified by an equivalence
class of pairs $(B,\mu)$ under the action of $\W$, where $B\in\t$
is a magnetic weight, and $\mu\in \t^*$ is an ordinary weight of
$\g$.

\subsection{S-duality}

Let us now discuss the action of the S-duality group on the set of Wilson-'t Hooft operators.
Recall that for $N=4$ super-Yang-Mills theory with a simply-laced simple Lie algebra $\g$ the duality group is conjectured to be $SL(2,\ZZ)$, while for non-simply laced Lie algebras it is a subgroup
of $SL(2,\ZZ)$ denoted $\Gamma_0(q)$, where $q=2$ for $\g=\so,\sp, F_4$ and $q=3$ for $\g=G_2$ (see
Refs.~\cite{GGPZ,doreyetal}). We remind
that the group $\Gamma_0(q)$ is a subgroup of $SL(2,\ZZ)$ consisting of the matrices of the form
$$
\begin{pmatrix} a & b\\ c & d\end{pmatrix},\quad a,b,c,d\in \ZZ,\ ad-bc=1,\ c\equiv 0\!\!\mod q.
$$
The group $SL(2,\ZZ)$ is generated by three elements $S,T,C,$
where $C=-1$ is central and the following relations hold:
\begin{equation}\label{sl2defrel}
C^2=1,\ S^2=C,\ (ST)^3=C.
\end{equation}
The group $\Gamma_0(q)$ is generated by $C,T,$ and $ST^qS$.

The element $C$ is represented by charge-conjugation. It reverses the sign of the Yang-Mills potential
$A$ and curvature $F$, so it acts on Wilson-'t Hooft operators by
$$
C: (\mu,B)\mapsto (-\mu,-B).
$$
(If the group $G$ admits only self-conjugate representations, $C$ is a trivial operation. In that case, $-1$
is in the Weyl group, and multiplying $(\mu,B)$ by $-1$ does not affect the equivalence class of the pair $(\mu,B)$.)

The element $T$ corresponds to the shift of the $\theta$-angle by
$2\pi$. At this stage it is important to fix a normalization of
the invariant metric on $\g$. We want the coefficient of $\theta$
in the action to take value $1$ on the minimal instanton. An
instanton with a minimal action is obtained by embedding the usual
$SU(2)$ instanton into an $SU(2)$ subgroup of $\tG$ associated
with a short coroot. For such an instanton, the topological charge
is
$$
\frac{1}{2}\,\langle H_\alpha, H_\alpha\rangle.
$$
Hence we choose the metric so that short coroots have length $\sqrt 2$. Having fixed the metric, we
define the field theory action to be
$$
S=\frac{1}{2 g^2}\int \langle F_{\mu\nu},F_{\mu\nu}\rangle-\frac{i\theta}{8\pi^2}\int \langle F,\wedge F\rangle.
$$
In the case of $G=SU(N)$ this normalization of $g^2$ and $\theta$ is equivalent to the following
standard one:
$$
S=\frac{1}{2g^2}\int \Tr( F_{\mu\nu} F_{\mu\nu})-\frac{i\theta}{8\pi^2}\int \Tr(F\wedge F),
$$
where $\Tr$ is the trace in the fundamental representation.

Having fixed a metric on $\g$, we get a natural isomorphism
$\ell:\t\ra \t^*$. For any element $a\in\t$ we let
$a^*=\ell(a)\in\t^*$. Similarly, for any $\rho\in \t^*$ we let
$\rho^*=\ell^{-1}(\rho)$ be the corresponding element of $\t$. We
claim that the T-transformation acts as follows:
\begin{equation}\label{Ttrans}
T: (\mu,B)\mapsto (\mu+B^*,B).
\end{equation}
This makes sense, because for any coroot $H_\alpha\in \t$ we have,
by Eq.~(\ref{alphadual}),
$$
B^*(H_\alpha)=H_\alpha^*(B)=\frac{1}{2} \langle H_\alpha,
H_\alpha\rangle\, \alpha(B).
$$
Since short coroots have been normalized to satisfy
$$
\langle H_\alpha, H_\alpha\rangle=2,
$$
and the length-squared of long coroots is an integral multiple of that for short coroots,
we see that $B^*(H_\alpha)$ is necessarily an integer, and therefore $\mu+B^*$ belongs to the
weight lattice. It is also easy to see that the map Eq.~(\ref{Ttrans}) commutes with the action
of the Weyl group $\W$, so we get a well-defined map on the set of orbits of $\W$.

Now let us show that shifting the $\theta$-angle by $2\pi$ induces
the map Eq.~(\ref{Ttrans}) on Wilson-'t Hooft operators. Recall
that $\mu\in\t^*$ specifies a representation $R_\mu$ of $\tG_B$.
The reductive Lie algebra $\g_B$ is a direct sum of a semi-simple Lie
algebra $\g_{ss}$ and an abelian Lie algebra $\g_{ab}=\t_{ab}$. The
transformation Eq.~(\ref{Ttrans}) modifies only the action of
$\g_{ab}$. Indeed, for any root $\alpha\in\Phi_B$ we have, by
Eqs.~(\ref{alphadual},\ref{phiBcond}),
$$
B^*(H_\alpha)=\frac{1}{2} \langle H_\alpha, H_\alpha\rangle\,
\alpha(B)=0.
$$
Thus on $\t_{ss}$ the weight $\mu+B^*$ agrees with $\mu$. The
$\mu$-dependent factor in the path-integral given by
Eq.~(\ref{wilsonR}) factorizes into a semi-simple piece and an
abelian piece. We want to show that shifting the $\theta$ by
$2\pi$ is equivalent to multiplying the abelian piece by
\begin{equation}\label{want}
\exp\left(i\int B^*(A_0(t,0)) dt\right).
\end{equation}
This is shown in the same way as in Section~\ref{sec:free}. We note that the topological term in the
action reduces to a boundary term of the form
$$
S_\theta=-\frac{i\theta}{4\pi^2}\int_{r=\eps} \langle A_0, B_i
n^i\rangle d^2\sigma dt,
$$
where $B_i=\frac{1}{2}\eps_{ijk}F_{jk}$ is the magnetic field, and
the integration is over an $S^2\times\RR\subset \RR^4$ given by
the equation $r=\eps$. Taking into account the behavior of $B_i$
for small $r$, performing the integral over $S^2,$ and setting
$\theta=2\pi$, the above expression becomes
$$
-i  \int \langle A_0 (t,0), B\rangle dt.
$$
Thus for $\theta=2\pi$ the exponential of $-S_\theta$ is precisely given by Eq.~(\ref{want}), as claimed.

Finally, we would like to determine how $S$ (or in the
non-simply-laced case, $ST^qS$) acts on the pairs $(\mu,B)$. Since
S-duality is still a conjecture, we have to guess the
transformation law. Guided by the analogy with the abelian case,
we propose that $S$ acts as follows:
$$
S: (\mu,B)\mapsto (-B^*,\mu^*).
$$
This transformation law has been previously considered in Ref.~\cite{doreyetal} in a somewhat
different context.
The same argument as above shows that $B^*(H_\alpha)$ is integral for all $\alpha\in\Phi$, so
$-B^*$ is a weight of $\g$. On the other hand, we have
\begin{equation}\label{scheck}
\alpha(\mu^*)=\mu(\alpha^*)=\frac{2\mu(H_\alpha)}{\langle H_\alpha, H_\alpha\rangle}.
\end{equation}
If $\g$ is simply-laced, all coroots have length-squared equal to $2$, so $\mu^*$ is
a magnetic weight. But for non-simply laced $\g$ we also have longer coroots, so
$\alpha(\mu^*)$ is not necessarily integral for all $\alpha\in\Phi$. Thus $S$ as defined above
is a well-defined operation on the set of Wilson-'t Hooft operators only in the simply-laced
case. It is easy to check that $C,T,S$ defined above satisfy the relations Eq.~(\ref{sl2defrel}).

In the non-simply-laced case we consider a transformation $ST^qS$ which acts as follows:
$$
ST^q S: (\mu,B)\mapsto (-\mu,q\mu^*-B).
$$
If we want this to be a legal transformation of Wilson-'t Hooft operators, then $q\alpha(\mu^*)$
must be an integer for all $\alpha\in\Phi$. According to Eq.~(\ref{scheck}), $\alpha(\mu^*)$ is an integer multiple of $1/2$ for $\g=\so,\sp,$ and $F_4$, and an integer multiple of $1/3$ for $\g=G_2$. Thus for
$\g=\so,\sp,$ and $F_4$ the largest possible duality group (among congruence subgroups of the
form $\Gamma_0(q)$) is $\Gamma_0(2)$, while for $\g=G_2$ it is $\Gamma_0(3)$. This agrees with
Refs.~\cite{GGPZ,doreyetal}.

\subsection{BPS Wilson-'t Hooft operators}

In the case of $N=4$ SYM theory it is natural to consider line operators which preserve some supersymmetry.
The simplest of these are $1/2$ BPS line operators. In the purely electric case, they are well-known:
$$
W^{BPS}_R=\Tr_R P\exp\left(\int \left(iA_0(t,0)+\phi(t,0)\right) dt\right),
$$
where $\phi$ is one of the real scalars in the $N=4$ multiplet,
and $R$ is an irreducible representation of $G$. They are
sometimes called Maldacena-Wilson operators because of their role
in AdS/CFT correspondence~\cite{sjr1,malda,sjr2}. The scaling weight of the
BPS Wilson operator at weak coupling is
\begin{equation}\label{scweightWBPS}
h\left(W^{BPS}_R\right)=\frac{g^2 c_2(R)}{96\pi^2}+O(g^4).
\end{equation}

Let us also define a BPS version of the 't Hooft operator. In
addition to fixed boundary conditions for the gauge field, as in
Eq.~(\ref{hooftbdrycond}), we impose a fixed boundary condition
for the scalar fields:
$$
\phi^a=\frac{\lambda^a}{r}+O(1), \ a=1,\ldots,6,
$$
where for each $a$ $\lambda^a$ is a covariantly constant section
of the adjoint bundle on $S^2\times\RR$. These sections are
determined by the BPS condition. To leading order in the gauge
coupling, we may simply require that the solution of the classical
equations of motion with the above asymptotics be BPS. This
implies that
$$
\lambda^a=\frac{1}{2}n^a B,
$$
where $n^a$ is a unit vector in $\RR^6$. Using $SO(6)$ R-symmetry,
we may always rotate $n^a$ so that only one of its components is
nonzero.

The general case of a BPS Wilson-'t Hooft operator is more
complicated and will not be treated here fully. We only note that
to leading order in $g^2$ one can simply neglect the ``Wilson''
part of the operator; then the behavior of the scalar field at
$r=0$ is the same as determined above.

\subsection{Scaling weights of Wilson-'t Hooft operators in $N=4$ super-Yang-Mills theory}\label{sec:dims}

In this subsection we compute the scaling weights of Wilson-'t
Hooft operators (both BPS and non-BPS) in $N=4$ super-Yang-Mills
theory at weak coupling. Since properties of purely electric line
operators (Wilson loops) are well-understood, we will focus on the
case when the GNO ``charge'' $B\in\Lambda_{mw}$ is nonzero.
Recalling the computations in Section~\ref{sec:free}, one can
easily see that to leading order in the $g^2$ expansion one can
neglect both the ``Wilson'' part of the operator, and the
$\theta$-term in the action. Therefore to this order it is
sufficient to consider 't Hooft operators and set $\theta=0$.

To evaluate the scaling weight to leading order in the
weak-coupling expansion, we can simply evaluate the classical
stress-energy tensor on the solution of classical equations of
motion with the desired asymptotics at $r=0$. This solution is
$$
A_\mu dx^\mu=\frac{B}{2}(1-\cos\theta) d\phi,\quad
\phi=\frac{\eta B}{2r}.
$$
Here $\eta=0$ corresponds to the ordinary 't Hooft operator (the S-dual of the ordinary Wilson
operator), while $\eta=1$ corresponds to the BPS 't Hooft operator. The bosonic part of the
classical stress-energy tensor is
\begin{multline*}
T_{\mu\nu}=2g^{-2}\Tr\left[D_\mu\phi\,D_\nu\phi-\frac{1}{2}\delta_{\mu\nu}(D\phi)^2
-\frac{1}{6}\left(D_\mu D_\nu-\delta_{\mu\nu}D^2\right)\phi^2\right]\\
+2g^{-2}\Tr\left[-F_{\mu\lambda}F_{\nu\lambda}+\frac{1}{4}\delta_{\mu\nu} F_{\lambda\rho} F_{\lambda\rho}\right].
\end{multline*}
One easily finds that the scaling weight is
\begin{equation}\label{scweightH}
h(H_{B,\eta})=\frac{1-\frac{\eta^2}{3}}{4g^2}\langle B,B\rangle +O(1).
\end{equation}

S-duality predicts that the scaling weight of the 't Hooft
operator at coupling $g$ is equal to the scaling weight of the
Wilson operator at coupling ${\hat g}=4\pi/g$. Our weak-coupling
results Eqs.~(\ref{scweightW},\ref{scweightWBPS}) and
Eq.~(\ref{scweightH}) show that this can be true only if the
scaling weights of both Wilson and 't Hooft operators (whether BPS
or not) receive higher-order corrections. Indeed, even the
group-theory dependence of their scaling weights is very different
at weak coupling. For example, for $G=SU(2)$ the scaling weight of
the Wilson operator in the representation of isospin $j$ goes like
$$
h\sim j(j+1),
$$
while the scaling weight of the 't Hooft operator of ``magnetic isospin'' $j$ goes like
$$
h\sim j^2.
$$

\section{Discussion and outlook}\label{sec:summ}

In this paper we have studied line operators in 4d gauge theories
which create electric and magnetic flux. In a free theory with
gauge group $U(1)$, such operators are classified by a pair of
integers, the electric and magnetic charges. Taking into account
the results of Goddard, Nuyts, and Olive~\cite{GNO}, one could
guess that in the nonabelian case Wilson-'t Hooft operators are
classified by a pair of irreducible representations, one for the
original gauge group $G$ and the other for its magnetic dual
$\hG$. We will denote the set of irreducible representations of
$G$ by ${\rm IRep}(G)$.

With a little more thought, however, one realizes that in a
nonabelian theory there should be some interaction between the
electric and magnetic representations. Our results show that such
an interaction, although present, has a very simple form: instead
of being labelled by a pair of irreducible representations,
Wilson-'t Hooft operators are labelled by an element of
$\Lambda_w\times\Lambda_{mw}$, modulo the Weyl group of $\g$.
There is an obvious map from this set to the set ${\rm
IRep}(G)\times {\rm IRep}(\hG)$, but this map is not injective. In
other words, there is more information in a pair of weights modulo
the action of the Weyl group than in the corresponding representation of
$G\times\hG$.

Let us illustrate this in the simple case $\g=\sl_2$. Both lattices
$\Lambda_w$ and $\Lambda_{mw}$ are one-dimensional in this case
and can be identified with $\ZZ$. Thus a weight (either electric
of magnetic) is simply an integer. The Weyl group is $\ZZ_2$, and
its nontrivial element acts by negating both integers. An integer
$m\in\ZZ$ corresponds to a representation of $SU(2)$ with isospin
$j=|m|/2$. We see that if both electric and magnetic weights are
nonzero, then there is precisely one more bit of information in
the pair of weights than in the corresponding pair of
representations. One can think of it as a ``mutual orientation''
of electric and magnetic representations.

This ``interaction'' between electric and magnetic representations
makes it possible to have an action of the S-duality group on the
set of Wilson-'t Hooft operators. Let us again illustrate our
point in the example of $\g=\sl_2$, where the S-duality group is
$SL(2,\ZZ)$. As explained above, one can index Wilson-'t Hooft
operators by a pair of isospins $j_e,j_m$ and an extra label
$\xi\in \{1,-1\}$. The T and S transformations act as follows:
$$
T: (j_e,j_m,\xi)\mapsto (|j_e + \xi j_m|,j_m,{\rm sign}(\xi j_e+j_m)),\quad S:
(j_e,j_m,\xi)\mapsto (j_m,j_e,-\xi).
$$
In contrast, no nontrivial $SL(2,\ZZ)$ action is possible on the
set of pairs of isospins.

In this paper we analyzed the action of S-duality on the Wilson-'t
Hooft operators in $N=4$ $d=4$ SYM theory. It would be interesting
to understand the action of other proposed dualities on line
operators in $d=4$ supersymmetric gauge theories. For example, it would be interesting
to understand the action of dualities on line operators in finite $N=2$ quiver theories.
The conjectured duality group in this case is much more complicated than for $N=4$ SYM: it is the fundamental group of the moduli space of flat irreducible connections on $T^2$, where the gauge group is simply-laced and determined by the type of the quiver~\cite{KMV,Dquiver}.
One could also ask how the Wilson loop operator in $N=1$ super-QCD
transforms under Seiberg's duality~\cite{Seiberg}. Unlike in the $N=4$ case, the
physical origin of Seiberg's duality is obscure, and it is not
clear whether Wilson operators are mapped to t' Hooft operators. A
hint that this may indeed be so is provided by a work of
M.~Strassler~\cite{Strassler}, who argued that in $N=1$ SUSY QCD with
gauge group $SO(N)$ and vector matter Seiberg's duality maps
Wilson operators corresponding to the spinor representation to
line operators carrying nontrivial 't Hooft magnetic flux.

Another possible line of investigation is to study line operators
in 3d theories. For example, one can ``uplift'' twist operators
for free fermions and bosons in 2d to line operators creating
topological disorder in the corresponding free 3d theories. A more
nontrivial example is the ``barbed-wire'' line operator in the 3d
Ising model defined by Dotsenko and Polyakov~\cite{DP}. Recall that
the 3d Ising model is related by Kramers-Wannier duality to a
$\ZZ_2$ gauge theory. The most obvious line operator in  this
theory is the Wilson operator. The ``barbed-wire'' operator is obtained
by decorating the Wilson operator with the Ising spin
operators. Dotsenko and Polyakov showed that the
``barbed-wire'' operator satisfies a linear equation, which looks like a
loop-space generalization of the Dirac equation. On the basis of
this observation, they conjectured that the 3d Ising model may be
integrable when expressed in terms of the ``barbed-wire'' operators. It
would be interesting to test this conjecture by computing
correlators of the ``barbed-wire'' operators with themselves and with
local operators.

\appendix
\section{Lie algebra facts and conventions}\label{appendix}

In this appendix we record some basic definitions and conventions
pertaining to compact simple Lie algebras and Lie groups. A
standard reference on these matters is Ref.~\cite{Humph}. Let $\g$
be such a Lie algebra. It has an $Ad$-invariant metric, which is
unique up to a rescaling. We will not fix any particular
normalization of the invariant metric on $\g$, and therefore will
not identify $\g$ and $\g^*$. Let $\t$ be a Cartan subalgebra of
$\g$ (i.e. a maximal abelian subalgebra of $\g$), and let
$\Phi\in\t^*$ be the set of roots of $\g$. This means that
$\g_\CC$ decomposes as
$$
\g_\CC=\t_\CC \op \bigoplus_{\alpha\in\Phi} V_\alpha ,
$$
such that for any $H\in \t$ and any $X\in V_\alpha$ we have
$$
[H,X]=\alpha(H)X.
$$
The subspaces $V_\alpha$ are called root spaces; they can be shown
to be one-dimensional. The span of $\Phi$ is the whole $\t^*$.

It is always possible to choose a basis vector $E_\alpha$ for each
$V_\alpha$ and a vector $H_\alpha\in \t$ for each $\alpha\in\Phi$
so that
$$
[E_\alpha,E_{-\alpha}]=H_\alpha,\quad [H_\alpha,E_{\pm\alpha}]=\pm
2E_{\pm\alpha}.
$$
The vector $E_\alpha\in V_\alpha$ is called the root vector
corresponding to the root $\alpha\in \t^*$, while $H_\alpha\in\t$
is called the coroot corresponding to the root $\alpha$. One can
show that
$$
\alpha(H_\beta)\in\ZZ, \quad \forall \alpha,\beta\in\Phi.
$$
The set of coroots spans $\t$.

Using the restriction of the invariant metric to $\t$, we can
associate a vector $\alpha^*\in\t$ to each root $\alpha\in\t^*$.
One can show that $\alpha^*$ is proportional to $H_\alpha$, while
their norms are related by
$$
\langle \alpha^*,\alpha^*\rangle \cdot\langle H_\alpha,
H_\alpha\rangle=4.
$$
This relation is independent of the particular normalization of
the invariant metric. We will use this relation in the following
form:
\begin{equation}\label{alphadual}
\alpha^*=\frac{2 H_\alpha}{\langle H_\alpha, H_\alpha\rangle},
\quad H_\alpha^*=\frac{1}{2}\alpha \langle H_\alpha,
H_\alpha\rangle .
\end{equation}

Choosing a particular Cartan subalgebra breaks the gauge group
down to a subgroup. The residual gauge transformations acting
nontrivially on $\t$ form a finite group $\W$ called the Weyl
group of $\g$. It consists of linear transformations $w_\alpha$,
$\alpha\in\Phi$, of the form
$$
w_\alpha(H)=H-\alpha(H) H_\alpha,\ \forall H\in\t.
$$
These linear transformations are called Weyl reflections. The set
of coroots is invariant with respect to the action of $\W$. The
Weyl group also acts on the dual space $\t^*$ as follows:
$$
w_\alpha(f)=f-f(H_\alpha)\alpha, \ \forall f\in\t^*.
$$
The set of roots $\Phi$ is invariant with respect to this action.

The roots of $\g$ span a lattice $\Lambda_r$ in $\t^*$ called the
root lattice of $\g$. Similarly, the coroots of $\g$ span a
lattice $\Lambda_{cr}$ in $\t$, called the coroot lattice. The
dual of the coroot lattice is a lattice $\Lambda_w$ in $\t^*$
defined by the condition
$$
f\in\Lambda_w\iff f(H_\alpha)\in\ZZ,\ \forall \alpha\in\Phi.
$$
This lattice is called the weight lattice of $\g$, and its
elements are called weights of $\g$. It is easy to see that the
root lattice $\Lambda_r$ is a sublattice of $\Lambda_w$. One can
also show  that the quotient lattice $\Lambda_w/\Lambda_r$ is
isomorphic to the center of $\tG$, the unique simply connected
compact Lie group with Lie algebra $\g$.

Dually, in $\t$ we have a lattice $\Lambda_{mw}$ defined by the
condition
$$
H\in\Lambda_{mw}\iff \alpha(H)\in \ZZ,\ \forall \alpha\in\Phi.
$$
This lattice is dual to the root lattice $\Lambda_r$ and is called
the lattice of magnetic weights of $\g$ in the main text. The
lattice $\Lambda_{cr}$ is a sublattice of $\Lambda_{mw}$, and
their quotient is again the center of $\tG$.

Let $G$ be a compact Lie group with Lie algebra $\g$. The kernel
of the exponential mapping
$$
\t\ra G,\quad H\mapsto \exp(2\pi i H)
$$
is a yet another lattice in $\t$, which we call $\Gamma_G$. One
has the inclusions
$$
\Lambda_{cr}\subset \Gamma_G\subset \Lambda_{mw}.
$$
The center and the fundamental group of $G$ can be determined as
follows:
$$
Z(G)=\Lambda_{mw}/\Gamma_G,\quad \pi_1(G)=\Gamma_G/\Lambda_{cr}.
$$
The dual of $\Gamma_G$ is a lattice in $\t^*$ known as the weight
lattice of $G$. We will denote it $\Gamma_G^*$. Obviously, we have
inclusions
$$
\Lambda_r\subset \Gamma^*_G\subset \Lambda_w
$$
and group isomorphisms
$$
Z(G)=\Gamma_G^*/\Lambda_r,\quad \pi_1(G)=\Lambda_w/\Gamma_G^*.
$$

A compact simple Lie algebra is called simply-laced if all its roots (i.e. all
elements of $\Phi$) have the same length, and non-simply-laced
otherwise. The simply-laced Lie algebras are $\sl_N$ and
$\so_{2N}$ series and the exceptional Lie algebras $E_6,E_7,E_8$.
In the non-simply-laced case, the roots can have only two
different lengths, so one can meaningfully talk about short roots
and long roots. The ratio of lengths-squared of long and short
roots is either $2$ (for Lie algebras $\sp_N,\so_{2N+1},$ and
$F_4$) or $3$ (for the exceptional Lie algebra $G_2$).

A compact simple Lie algebra can be reconstructed from its set of
roots $\Phi\in\t^*$ (provided that the invariant metric is also
specified). A Lie algebra $\hg$ is called the magnetic dual of
$\g$ if its set of roots coincides with the set of coroots of
$\g$. Each compact simple Lie algebra has a magnetic dual;
applying the duality procedure twice gives back the Lie algebra
one has started from. We also have the notion of a magnetic dual
group $\hG$: if $G$ is a compact simple Lie group with Lie algebra
$\g$ and the kernel of the exponential mapping
$\Gamma_G\subset\t$, then the magnetic dual group $\hG$ is
uniquely defined by requiring that its Lie algebra be $\hg$, and
its kernel of the exponential mapping $\Gamma_{\hG}\subset
\hatt=\t^*$ be the weight lattice $\Gamma_G^*$. In particular, if
$G=\tG$, then the magnetic dual group has
$\Gamma_{\hG}=\Lambda_w$. By the above definitions, the lattice of
magnetic weights for $\hg$ is the weight lattice $\Lambda_w$ for
$\g$, therefore the magnetic dual group $\hG$ has a trivial
center. In the main text such a group was denoted $\hG_0$.

\section*{Acknowledgments}

I would like to thank Andrei Mikhailov for helpful discussions.
This work was supported in part by the DOE grant DE-FG03-92-ER40701.

\end{document}